%% file: preprint-template.tex
\RequirePackage[2020-02-02]{latexrelease}
\documentclass[twocolumn, switch]{article} 

\usepackage{amsmath, amsthm, amssymb, amsfonts}

\usepackage[numbers,square]{natbib}
\usepackage[
 left=20mm,
 top=25mm,
 right=20mm,
 bottom=25mm,
 headsep=12.5mm,
 headheight=14.5pt, 
 heightrounded
]{geometry}

\usepackage[utf8]{inputenc}	
\usepackage[T1]{fontenc}	
\usepackage{xcolor}		
\usepackage[colorlinks = true,
            linkcolor = purple,
            urlcolor  = blue,
            citecolor = cyan,
            anchorcolor = black]{hyperref}	
\usepackage{booktabs} 		
\usepackage{nicefrac}		
\usepackage{microtype}		
\usepackage{lineno}		
\usepackage{float}			

\usepackage{tabularx}
\usepackage{adjustbox}
\usepackage{subcaption}
\usepackage{lipsum}		
\usepackage{url}

\usepackage{breakurl}

\usepackage{newfloat}
\DeclareFloatingEnvironment[name={Supplementary Figure}]{suppfigure}
\usepackage{sidecap}
\sidecaptionvpos{figure}{c}

\usepackage{titlesec}
\titlespacing\section{0pt}{12pt plus 3pt minus 3pt}{1pt plus 1pt minus 1pt}
\titlespacing\subsection{0pt}{10pt plus 3pt minus 3pt}{1pt plus 1pt minus 1pt}
\titlespacing\subsubsection{0pt}{8pt plus 3pt minus 3pt}{1pt plus 1pt minus 1pt}

\usepackage{tikz,xcolor,hyperref}

\definecolor{lime}{HTML}{A6CE39}
\DeclareRobustCommand{\orcidicon}{
	\begin{tikzpicture}
	\draw[lime, fill=lime] (0,0) 
	circle [radius=0.16] 
	node[white] {{\fontfamily{qag}\selectfont \tiny ID}};
	\draw[white, fill=white] (-0.0625,0.095) 
	circle [radius=0.007];
	\end{tikzpicture}
	\hspace{-2mm}
}
\foreach \x in {A, ..., Z}{\expandafter\xdef\csname orcid\x\endcsname{\noexpand\href{https://orcid.org/\csname orcidauthor\x\endcsname}
			{\noexpand\orcidicon}}
}

\title{Individual and Collective Performance Deteriorate in a New Team: A Case Study of {\itshape CS:GO} Tournaments}

\usepackage{xwatermark}
\newwatermark[firstpage,color=gray!90,angle=0,scale=0.28, xpos=0in,ypos=-5in]{*correspondence: \texttt{gmuric@isi.edu}}

\usepackage{authblk}

\author[1]{Weiwei Zhang}
\author[2,*]{Goran Muric\orcidB{}}
\author[1,2,3]{Emilio Ferrara\orcidC{}}

\affil[1]{Annenberg School of Communication, University of Southern California , Los Angeles, CA, US}
\affil[2]{Information Sciences Institute, University of Southern California , Marina Del Rey, CA, US}
\affil[3]{Department of Computer Science, Viterbi School of Engineering, University of Southern California, Los Angeles, CA, US}

\begin{document}

\twocolumn[ 
\begin{@twocolumnfalse} 
  
\maketitle

\begin{abstract}
How does the team formation relates to team performance in professional video game playing? This study examined one aspect of group dynamics - team switching - and aims to answer how changing a team affects individual and collective performance in eSports tournaments. In this study we test the hypothesis that switching teams can be detrimental to individual and team performance both in short term and in a long run. We collected data from professional tournaments of a popular first-person shooter game {\itshape Counter-Strike: Global Offensive (CS:GO)} and perform two natural experiments. We found that the player's performance was inversely correlated with the number of teams a player had joined. After a player switched to a new team, both the individual and the collective performance dropped initially, and then slowly recovered. 
The findings in this study can provide insights for understanding group dynamics in eSports team play and eventually emphasize the importance of team cohesion in facilitating team collaboration, coordination, and knowledge sharing in teamwork in general. 
\end{abstract}
\vspace{0.35cm}

  \end{@twocolumnfalse} 
] 


\section{Introduction}
The eSports industry emerged since the 1980s~\citep{borowy2013pioneering}, and it has currently shared a significant proportion of the entertainment industry~\citep{reitman2020esports}. The estimated global revenue in eSports has surged from 776 million USD in 2018 to over 950 million USD in 2020~\citep{rietkerk_2020}. The market share of eSports is expected to keep expanding, with the global revenue estimated to soar to 1.6 billion USD in 2023~\citep{rietkerk_2020}. The population of eSports viewers keeps increasing by more than 10 percent annually: in 2020, it has been estimated that 646 million viewers watch eSports, and 295 million of them were labeled as enthusiasts~\citep{roundhill_2020}. 

Various professional tournaments have found their place into the eSports industry~\citep{sledge_2020}, and this motivates many aspiring and young video game players to play games as a profession~\citep{kim2015stage, kocadaug2019investigating, salo2017career, meng2020understanding}. In response to this popularity, multiple disciplines have paid serious attention on eSports research. As evident by a prior study~\citep{reitman2020esports}, there had been 150 articles published on eSports from 2002 through March 2018 from different disciplines such as sociology, sports science, business, and other fields. However, papers that were concentrating on team play in eSports mostly evaluated how training and coaching affect team performance~\citep{kari2016athletes, nagorsky2020structure}, instead of examining team dynamics and group composition, except a few~\citep{parshakov2018diversity, parshakov2018determinants, parshakov2019economics}. On the other hand, though many existing studies have investigated what impacts video game playing in ad-hoc teams~\citep{kim2016proficiency, sapienza2017performance, sapienza2018individual, wang2019personality}, their findings might not be generalized to the contexts of professional tournaments. 

To fill this gap, this study focused on one aspect of group dynamics in eSports - team switching - and investigated how switching teams affects individual players' and their teams' performance.  

\subsection{eSports as a profession}
Professional video game players have been found to have different motivations and playing patterns than general video game players. Previous studies show that eSports athletes are much more enthusiastic about attending competitions, and they desire a stronger sense of community belonging than those who play video games as a hobby~\citep{martonvcik2015sports}. For professional eSports athletes, video gaming is an activity of both leisure and work, and beyond that, also part of their life goal~\citep{martonvcik2015sports, seo2016professionalized}. Besides holding different game-play motivations than ordinary video game players, eSports professionals are more physically and psychologically competent. eSports players devote their time to training physical conditions~\citep{kari2016athletes} and endeavoring to equip themselves with high-level mental capacity and cognitive alertness to compete in matches~\citep{banyai2019psychology,himmelstein2017exploration, meng2020understanding, khromov2019esports, parshakov2019economics}. Other than that, because of eSports' digital nature, players are also required to command technological proficiency to respond fast and accurately in the game~\citep{freeman2017esports,nagorsky2020structure, witkowski2012digital}.

Though the eSports industry has been believed to share similarities with traditional sports in terms of organizing, institutionalization, and competitive nature~\citep{banyai2019psychology,hallmann2018esports, jenny2017virtual, kari2016athletes, nagorsky2020structure, thiel2018esport}, playing eSports as a profession could be much more demanding than traditional sports. Given that the eSports industry is still young, new orders are emerging every day~\citep{meng2020understanding}. Players not only have to deal with the stress of being watched by a large audience~\citep{banyai2019psychology, matsui2020does}, but also the stress to maintain the complex relationships between  multiple stakeholders~\citep{meng2020understanding}. And because of the stereotypes over video game playing, eSports players also have to be careful to manage their social circles outside the career~\citep{salo2017career}. 

Besides the obstacles mentioned above, the career span of eSports players is short, and their skills are not likely to be transferrable to other fields, which makes their career security extremely fragile~\citep{kim2015stage, meng2020understanding, salo2017career}. Therefore, it is crucial for eSports players and team sponsors to optimize teamplay strategies that enable players to succeed before they retire from the tenuous and short-lived career. 

\subsection{Team play in eSports}
ESports is a format of computer-supported collaborative work (CSCW) where team members collaborate in a virtually connected team~\citep{freeman2017esports}.
In virtual teams, social congruence, mutual trust, and shared goals are the key factors that facilitate knowledge sharing and team performance~\citep{baruch2012all}. As found in ad-hoc teams in \textit{League of Legends} (\textit{LoL}), besides technical proficiency, team coordination is also critical to the success and the enjoyment of team play~\citep{kim2016proficiency}. Based on another previous study on ad-hoc teams in \textit{Lol}, players have to communicate frequently in games to coordinate teamwork and facilitate collective performance~\citep{kou2014playing}. To achieve team congruence, video game players are inclined to team up with people they are familiar with, but not necessarily competent or alike, when they are identifying temporary teammates~\citep{alhazmi2017empirical}. 

In professional teams in eSports, players have to have an agreeing personality and the willingness to communicate with teammates to achieve team cohesion~\citep{freeman2019understanding}. To demonstrate the potential of being a good teammate, eSports players should be physically and socially competent~\citep{benefield2016virtual, freeman2017esports, kim2016proficiency, nagorsky2020structure}. According to a previous study, eSports athletes tend to use social criteria to identify players who are likely to be congruent with the team environment as the potential new teammates to be recruited~\citep{freeman2019understanding}. 

In virtual teams, mutual trust also plays a crucial role in facilitating coordination, information collaboration, and knowledge sharing~\citep{choi2019mechanism, penarroja2015team, pinjani2013trust}. Mutual trust promotes a team environment that fosters collaboration~\citep{barczak2010antecedents}. In eSports teams, mutual trust among team members is even more critical than teams in other organizational settings. On one hand, conflicts and competitions have been seen widely in eSports teams, especially when team activities are tightly connected with business profits~\citep{lipovaya2018coordination}. On the other hand, in eSports competitions, players should feel comfortable to make decisions fast in a highly competitive and stressful environment~\citep{freeman2017esports}. Every team member is vital to success, given that each team member has their specialization in games. Team members should fit well with each other to in order to achieve efficient coordination~\citep{lipovaya2018coordination}. 

As in traditional team sports~\citep{van2020team}, we believe sharing team-level achievement goals among team members is also supposed to affect how players perform in eSports competitions. As evidenced by \citep{mukherjee2019prior}, teams with prior shared success are more likely to score higher in sports and eSports matches. This is understandable given that a group with shared goal is easier to coordinate in training than a random assembled mass. 

Team building is an effective way of establishing both cohesion, mutual trust and shared goals among virtual team members, as traditional sports team~\citep{carron2002cohesion}, and it has been demonstrated to be an effective way of enhancing both personal and team progression~\citep{breuer2016does, carron2002cohesion, rusman2010fostering}. And this is a long-term process that is not likely to take effect as a new member just joined a team. 

These findings all suggest that when a team lacks unity, mutual trust, and prior shared experiences, it is unlikely to help the player and the team achieve good performance and progression. Therefore, based on previous research, we believed that just after a player joined a new team, the player cannot quickly cultivate some of these qualities among other team members, which is predicted to hamper essential components in group processes such as information gathering and team learning and eventually lead to bad performance.

\subsection{The present study}
Based on the literature mentioned above, we hypothesized that a player joining a new team will negatively affect the performance of both player and the team they joined, at least in the short run. We predicted that the performance of the player could be partially attributed to the number of teams a player used to play in.

For this study, we collected data from professional tournaments of {\itshape Counter-Strike: Global Offensive} ({\itshape CS: GO}), one of the most popular multiplayer first-person shooter games, and made several natural experiment designs with mixed-effects regression models to seek the answer to the following question: {\itshape How does switching teams affect individual and collective performance?}

We approached this question in the following two ways: 1) we evaluated how the number of teams players had in the past affects their individual performance, and 2) within every instance of team switching, we inspected how the individual and the teams' performance change in the short and the long run. We found an inverse relation between the number of teams a player has joined and the player's performance measured by several performance indicators. The players who switched many teams tend to have lower performance in average than the players who stick with few teams. Additionally, when investigating the immediate and long-term effects of a single team switch, we found that right after a player switched to a new team, both the player and teams cast less damage and have lower survival rates, which eventually harms their overall rating and the probability of winning; in the long run, both the individual and the collective performance slowly recover.

We believed the present study has the following contributions: 1) the study provides insights that can illuminate team play strategies for eSports professional athletes by generating empirical findings over a large-scale data set; 2) the study also sheds light on computer-mediated collaboration in general which is operated in environment that demands intensive activities and high-level coordination; and 3) the study bridges the gap of eSports research implemented on behavioral-log data by applying computational and rigorous statistical methods to understand the effects of switching teams~\citep{reitman2020esports}. 

\section{Materials and Methods}\label{sec:matandmethods}
\subsection{About {\itshape CS: GO}}
Released in August 2011, \textit{Counter Strike: Global Offensive}, abbreviated as \textit{CS:GO} is the fourth game in the \textit{Counter Strike} series and one of the most popular multiplayer first-person shooter video games. In \textit{CS:GO}, two teams fight against each other, and each team takes the side of either Terrorists or Counter-Terrorists. Terrorists will be assigned a specific task in each round, and they have to either plant the bomb or defend the hostages\footnote{The game mode with a bomb is usually played in most professional tournaments.}, while the Counter-Terrorists have to prevent the task of Terrorists from being completed. The game is played in rounds and each round ends either when all players in a team are dead or when Terrorists have completed their task. In professional tournaments, each match usually consists of 30 rounds. The two teams switch sides after 15 rounds, and the team that wins at least 16 rounds out of 30 wins the match.

In 2015, \textit{CS:GO} was the second most played online video games on Twitch, following the \textit{League of Legends}. As of 2018, \textit{CS:GO} generated 414 million USD worldwide and was one of the most profitable PC and console games~\citep{scarborough_2019}, estimated to have 606,850 concurrent players daily on average.~\footnote{https://steamcharts.com/app/730} 

Professional tournaments of \textit{CS:GO} have attracted enthusiastic fans worldwide. In the tournament \textit{IEM Katowice 2020} that was on the air from January 3, 2020, to February 24, 2020, there were 279,453 viewers on average every minute, with more than 1 million viewers at the peak.~\footnote{https://escharts.com/tournaments/csgo/iem-katowice-2020-csgo} 

\subsection{Data}
We used the data of matches in the \textit{CS:GO} tournaments that took place between 2012 and 2020. The data has been collected from the HLTV.org, one of the most widely used websites that covers statistics of professional tournaments in \textit{CS:GO}.~\footnote{https://www.hltv.org/} We collected information about all the matches, teams, players, and tournaments from the website. The data collection process took place from February 1, 2020 to March 18, 2020 using the scraping tools in Python.

The earliest match in our dataset dates back to September 13, 2012 and the most recent match was completed on March 16, 2020. We extracted 76,693 matches in total and among those we had 76,645 matches from which we were able to obtain complete information about all the players in both teams. We further removed matches with more or less than 5 players in either of the two teams, which is irregular and could be attributed to the errors in the website. After the initial data cleaning, the data on 76,086 matches in 3,658 tournaments remained with 760,860 player-match combinations. From these matches, we identified 4,824 distinct teams and 13,656 distinct players. For each match, we obtained the statistics for different performance indicators of each player in each match, which we describe more in details in \nameref{sec:measures}. 

In Table~\ref{tab:distributions} we presented the annual distribution of matches and players' debuts, where the player's debut is the time when the specific player appeared for the first time in any of the tournaments. You can see that the number of matches and the number of players are not equally distributed over the years. 

    \begin{table}
        \caption{Distributions of matches and players' debut by year.}
        \label{tab:distributions}
        \begin{adjustbox}{width=1\linewidth}
        \begin{tabular}{rllllllllll} 
        \toprule
        {\bf Start year} & 2012&2013&2014&2015&2016&2017&2018&2019&2020& {\it Total}\\
        \midrule
        {\bf \# Matches} 
        &463 &2,306 &2,422 
        &7,251 &11,234 &16,799 
        &17,901 &15,194 &2,516
        & 76,086\\
        (\%) &(0.6) &(3.0) &(3.2) &(9.5) &(14.8) &(22.1)
        &(23.5) &(20.0) &(3.3)\\
        {\bf \# Player debut} 
        &467 &926 &662 
        &1,995 &2,672 &3,131 
        &2,232 &1,427 &144 &13,656\\
        (\%) &(3.4) &(6.8) &(4.8) 
        &(14.6) &(19.6) &(22.9) &(16.3) &(10.4) &(1.1)\\
        \bottomrule
        \end{tabular}
        \end{adjustbox}
    \end{table}

\subsection{Measures}\label{sec:measures}
We had information of each player in all the matches they competed during the period mentioned above. We used the obtained measures to compute individual player and team performances as dependent variables. 

\subsubsection{Dependent variables}
Dependent variables in this study were made from performance indicators provided by HLTV. In our study we made a distinction between the \textit{Individual performance} and the \textit{Team performance}. 

\textbf{Individual performance.} We obtained the performance indicators that capture specific skill-level information about players in each match including: 1) Kills (K) - The number of times a player made a kill; 2) Head shots (HS) - The number of specific types of precision kills; 3) Assists (A) - The number of times a player participated in the kill by damaging the opponent; 4) Flash assists (F) - The number of times a player successfully flashed~\footnote{A flashbang is the non-lethal grenade that temporarily blinds anybody within its concussive blast} an enemy which then gets killed by another player; 5) Deaths (D) - The number of times a player got killed by the others. Additionally, we used the derived variables that measure the overall performance of players: 6) KAST - The fraction of rounds in which the player either had a kill, assist, survived or was traded; 7) K-D Diff - The difference between kills and deaths; 8) FK Diff - The difference between the times the player was the first one to kill and the first one to die; 9) ADR - The average damage a player deals per round; 10) Rating - The overall rating based one all performance indicator computed by HLTV. 

   \begin{table}
        \caption{Descriptive statistics of performance indicators (N = 760,860)}
        \label{tab:stats}
        \centering 
        \begin{center}
        \begin{adjustbox}{width=\linewidth}
        \begin{tabular}{rllllllll} 
        \toprule
        {\bf Variable} &Mean &SD &Min &25\% &50\% &75\% &Max &missing\\
        \midrule
        Kills (K) &17.6 &6.35 &0 &13 &17 &22 &79 &0\\
        Head shots (HS) &7.9 &3.85 &0 &5 &8 &10 &39 &0\\
        Assists (A) &4.1 &2.44 &0 &2 &4 &6 &25 &0\\
        Flash assists (F) &1.0 &1.3 &0 &0 &1 &2 &15 &24,820*\\
        Deaths (D) &17.7 &4.81 &0 &15 &18 &20 &69 &0\\
        KAST &69.2 &12.38 &5.6 &61.5 &70.0 &77.8 &100 &12,840**\\
        K-D Diff &-0.04 &7.25 &-28 &-5 &0 &5 &40 &0\\
        FK Diff &-0.00 &2.44 &-12 &-2 &0 &2 &15 &0\\
        ADR &74.8 &18.88 
        &1.0 &61.9 &73.8 &86.6 &206.7 &12,850**\\
        Rating &1.05 &0.35 &0.00 &0.81 &1.03 &1.26 &3.45 &0\\
        \bottomrule
        \multicolumn{9}{l}{* Not reported on HLTV}\\
        \multicolumn{9}{l}{** Missing for all matches before 2015}\\
        \end{tabular}
        \end{adjustbox}
        \end{center}
    \end{table}

It is noteworthy that HLTV included \textit{KAST} and \textit{ADR} as the performance indicators after 2015, and there were 24,820 matches without \textit{Flash assists} information in our dataset. And \textit{Rating} was calculated based on all the available indicators at the time of the matches. 

Table~\ref{tab:stats} shows the descriptive statistics of performance indicators on the player-match level, where each player is likely to have multiple data points for each performance indicator. The summary statistics were calculated from the 760,860 distinct player-match combinations obtained from the 76,086 matches. 

For all the performance indicators (except {\itshape Flash assists}), we were able to identify a bell-shaped distribution that did not exhibit outliers. Pairwise correlations of these indicators at the player-match level were presented at Table~\ref{tab:corr1}. The low values of most pairwise correlations suggest that these indicators capture different aspects of players' performance. 

To measure a player's \textit{Avg. performance} up to the current match, we sequentially computed the average of each indicators of a player in previous matches up to the current match.
    
    \begin{table}
        \caption{Pearson correlations of performance indicators for individual players (N = 760,860). The table only shows correlations of complete pairwise observations.}
        \label{tab:corr1}
        \centering 
        \begin{center}
        \begin{adjustbox}{width=\linewidth}
        \begin{tabular}{r r ccccccccc} 
        \toprule
        Variable & &1 &2 &3 &4 &5 &6 &7 &8 &9\\
        \midrule
        Kills (K) & 1 \\
        Head shots (HS)& 2 &0.69\\
        Assists (A)& 3 &0.23 &0.17\\
        Flash assists (F) & 4& 0.05 &-0.01 &0.54\\
        Deaths (D) & 5 & 0.18 &0.15 &0.19 &-0.05\\
        KAST & 6 & 0.50 &0.35 &0.29 &0.11 &-0.42\\
        K-D Diff & 7&0.76 &0.51 &0.07 &0.08 &-0.51 &0.72\\
        FK Diff & 8&0.41 &0.25 &0.05 &0.05 &-0.22 &0.42 &0.50\\
        ADR & 9&0.72 &0.57 &0.25 &0.01 &-0.24 &0.60 &0.80 &0.47\\
        Rating & 10&0.72 &0.51 &0.19 &0.08 &-0.46 &0.80 &0.94 &0.55 &0.88\\
        \bottomrule
        \end{tabular}
        \end{adjustbox}
        \end{center}
    \end{table}
    
\textbf{Team performance}. In the metadata we extracted from HLTV, we had 4,824 teams with distinct identifiers created by the website. However, we made new identifiers that better fit the definition of a team: when two consecutive matches played by a team defined by HLTV have overlaps in team members, we used a distinct identifier to label that team. Otherwise, when a team changes all five players, we regarded it as a new team and a new identifier was used to label the the team from the second match. After we relabeled the teams, we ended up with 6,051 teams in the dataset in total. When we referred to teams in the following sections, we used these updated team identifiers. 

With the new identifiers, we computed team performance at the current match as the average of each of the ten \textit{Individual performance} of the five team members in the team. Additionally we used a binary variable \textit{Win} to indicate if the team won or lost the match for each player-match combination. Similarly to the \textit{Individual performance}, from the second match of teams, we sequentially computed the average of each indicators of team performance in previous matches, to measure a team's \textit{Avg. performance} up to the current matches. 

\subsubsection{Independent variable}
\textbf{Match Index} indicates the index of the current match played by the player or the team. We used this variable as a proxy of the player's or the team's experience. The maximum number of matches ever played by a single player throughout their career is 1,815, and the maximum number of matches played by a team is 1,643. 
    
\subsubsection{Other information}
We also had additional variables to indicate other information about players:

\textbf{Continent} indicates the continent of a player based on the nationality. A player can be from any of the six continents: Europe (EU), Asia (AS), North America (NA), South America (SA), Oceania (OC), Africa (AF), with 5 players in the dataset missing this information. A distribution of players by continent is shown in Table~\ref{tab:nation}.

    \begin{table}
        \caption{Distributions of players' by continent (N = 13,656)}
        \label{tab:nation}
        \centering 
        \begin{center}
        \begin{adjustbox}{width=\linewidth}
        \begin{tabular}{cccccccc} 
        \toprule
        EU &AS &NA &SA &OC &AF &\#missing\\
        \midrule
        6,477 &3,005 &2,009 &1,246 &550 &364 &5\\
        \midrule
        47.4\% &22\% &14.7\% &9\% &4\% &2.6\% &~\\
        \bottomrule
        \end{tabular}
        \end{adjustbox}
        \end{center}
    \end{table}

\textbf{Prize} indicates the monetary prize in US dollars awarded in tournaments. Out of the 3,658 tournaments, the prize information was available for 1,606 tournaments (\textit{Mean} = 47,090 USD, \textit{SD} = 137,220 USD). For each match of a player, we computed the \textit{Avg. prize} of the tournaments that held their previous matches to indicate the level of tournaments the player mostly played at.

\textbf{Year} indicates the year when the player had their first match.

\subsection{Analysis}\label{sec:model}
We conducted two sets of analyses to examine how switching teams affects individual and team performance. First, only for players, by comparing those who had played in many teams and those who played only in a few teams, we evaluated how the frequency of switching teams affected individual performance in the long run. Then, for both individual players and teams, we compared their performance before and after the instance of team switching happened, in order to identify if there is a performance gap caused by the event in the short and long term. We created Coarsened Exact Matching designs (CEM) to complete the first part of analysis, and customized Regression Discontinuity designs for the second part. And we used mixed-effects models~\citep{pinheiro2006mixed} to examine the causal effect in both parts of analyses.

\subsubsection{Dividing players into Treatment and Control groups}
For the causal models used in our first part of analysis as introduced in \nameref{sec:cem}, we first needed to split the players into two groups: Treatment and Control. The players who frequently changed their teams before the current match were assigned to Treatment group, and the players who rarely or had never changed teams were assigned to Control group. Simply dividing the players into two groups based on the number of teams they played during their tenure was not an optimal solution, as the number of teams a player had participated before the current (observed) match is changing over time. To address this problem, we made different Treatment and Control group for every $n=10$ matches the players play. To divide the players we used two cut-off values on the distribution of number of teams the players played - players who fell in the top 20 percentile were assigned to the Treatment group while those who fell in the bottom 20 percentile were assigned to the Control group. To make the result robust, we dropped players who fell between the two cut-off values. Since only 931 out of 13,656 players had more than 200 matches in our data set, we only performed this splitting procedure from the 10$^{th}$ to the 200$^{th}$ match of players with an increment of 10. The cut-offs to separate the two groups at different \textit{Match index} are displayed in Table \ref{tab:cutoff}.

\subsubsection{Coarsened exact matching}\label{sec:cem}
To examine if the frequency of switching teams affected players' individual performance, we first performed a causal modelling technique called Coarsened Exact Matching (CEM). We started from the assumption that the frequency of switching teams heavily depends on numerous external factors, including the level of tournaments the player played in, the time a player started their career and a player's nationality. To preclude the effect of confounding variables on the causal inference, we improved the natural experiment design with CEM. While keeping the \textit{Match index} the same, we binned players by their exogenous factors, and evaluated the effect of team-switching frequency in bins by comparing players in the Treatment group and the Control group with similar covariates. We used CEM to create matched bins as it often outweighs other matching procedures in terms of statistical credibility when estimating the treatment effect in observational studies~\citep{iacus2012causal}.

    \begin{table}
        \caption{Cut-off values on the number of teams players joined to separate \textit{Treatment} and \textit{Control} groups at different \textit{Match index} (N = 13,656)}
        \label{tab:cutoff}
        \centering 
        \begin{center}
        \begin{adjustbox}{width=\linewidth}
        \begin{tabular}{rcccccccccc} 
        \toprule
        Match index &10 &20 &30 &40 &50
        &60 &70 &80 &90 &100\\
        \midrule
        C, T* &1,3 &2,4 &2,5 &2,6 
        &2,6 &2,6 &3,7 &3,7 &3,8 &3,8\\
        \midrule
        Match index &110 &120 &130 &140 &150
        &160 &170 &180 &190 &200\\
        \midrule
        C, T &3,8 &3,9 &3,9 &3,9 
        &4,9 &4,9 &4,9 &4,10 &4,10 &4,10\\
        \bottomrule
        \multicolumn{11}{l}{* Players who had participated in $\leq C$ teams at the current match were}\\
        \multicolumn{11}{l}{assigned to the \textit{Control} group, while those who had participated in $\geq T$}\\
        \multicolumn{11}{l}{teams were assigned to the \textit{Treatment} group.}
        \end{tabular}
        \end{adjustbox}
        \end{center}
    \end{table}

At certain \textit{Match index}, we used three external factors to match players in the Treatment and the Control groups into bins, including the \textit{Continent} of a player's nationality, \textit{Avg. prize} of the previous tournaments the player participated in, and the \textit{Year} the player had the first match. Then, on bins of matched players, we examined the treatment effect on player performance with mixed-effects models. 

Mixed-effects models extend generalized linear models by including both fixed-effects terms and random-effects terms in the model~\citep{pinheiro2006mixed}. Fixed-effects terms capture variation across different observation units, while random-effects capture variation within the same unit of observation. Since different bins of players are likely to have different baseline performance, mixed-effects models allow us to control for heterogeneity among players in the random-effects terms caused by exogenous variables and isolate the treatment effect in a fixed-effects term.

Here is an example to illustrate this procedure: to evaluate the performance of players at the $100^{th}$ match, we first select users who played at least 100 matches as our sample. Based on the distribution of the total number of teams they played in the previous 100 matches, we categorized players who had played in no more than 3 teams as the Control group, and players who had played in at least 8 teams as the Treatment group. Then, we matched players into bins with the three covariates by CEM. After that, we compared the average of the 10 performance indicators until the 100$^{th}$ match of Treatment vs. Control groups by mixed-effects models with a random intercept standing for each covariate bin. Bins with observations from only one group, either Treatment or Control, and not in other will be dropped during the modeling.

We built a set of the mixed-effects models called \textit{Model group I} using the formulation in Eq.~\ref{eq:baseline_model}, where \textit{Performance} can be the average value of any of the indicators in \textit{Individual performance}, depending on the variable we model, until the current match. For instance, to model the number of kills, the \textit{Performance} variable would be the average of \textit{Kills} a player made in the previous matches. 

	\begin{equation}
	\label{eq:baseline_model}
    \begin{aligned}
	Performance \sim  1 & + Treatment:Yes\\
	               & + (1\vert Covariate\ bin)
	\end{aligned}
    \end{equation}

\subsubsection{Regression Discontinuity Design}\label{sec:rdmodel}
We created a customized Regression Discontinuity Design (RDD) model to answer the following questions: 1) after a player switched teams, how did the \textit{player's} performance change?; and 2) after a team changed one player, how did the \textit{team's} performance change? We built on the idea of regression discontinuity design (RDD) to examine this problem. RDD is used here to test the hypothesis that after each time of switching team, both the baseline performance and the progression rates of players and teams change.

RDD is a method used to identify causal relationship in natural experiment designs where the treatment is associated with one of the time-dependent independent variables~\citep{hahn2001identification}. A cutoff value is assigned before or after the treatment is implemented and the causal relationship is usually formulated as:

\begin{equation}
\centering
\label{eq:rdd}
Y = \alpha + X\beta + T\beta_t + TX\beta_{tx} + \epsilon   
\end{equation}

where $X \in \mathcal{R}^N$ is a vector of discrete or continuous time-dependent variable and $T$ is a vector of binary indicators with an entry $t_i$ equal to 1 when $x_i$ is greater than some cutoff $\lambda$. In this way, $\beta_t \in \mathcal{R}$ estimates how the treatment variable affects the baseline of the outcome variable $Y$, and $\beta_{tx}$ estimates how the treatment variable affects the trend of $X$ on $Y$. Within this study's setting, every single time of switching teams can be seen as a treatment. 

\begin{figure}
    \centering
    \begin{subfigure}{\linewidth}
        \includegraphics[width=\linewidth]{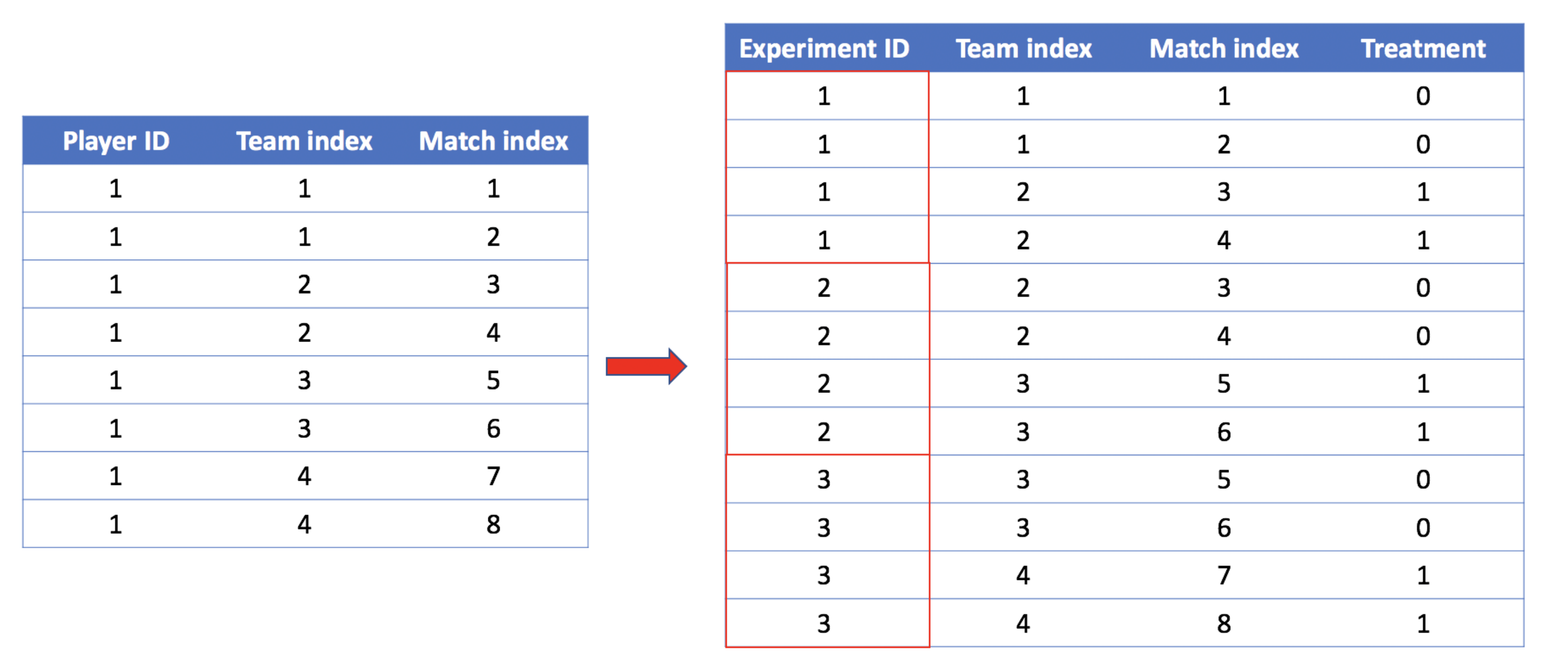}
        \caption{Players}
        \label{fig:quasidesign_player}
    \end{subfigure}
    \begin{subfigure}{\linewidth}
        \includegraphics[width=\linewidth]{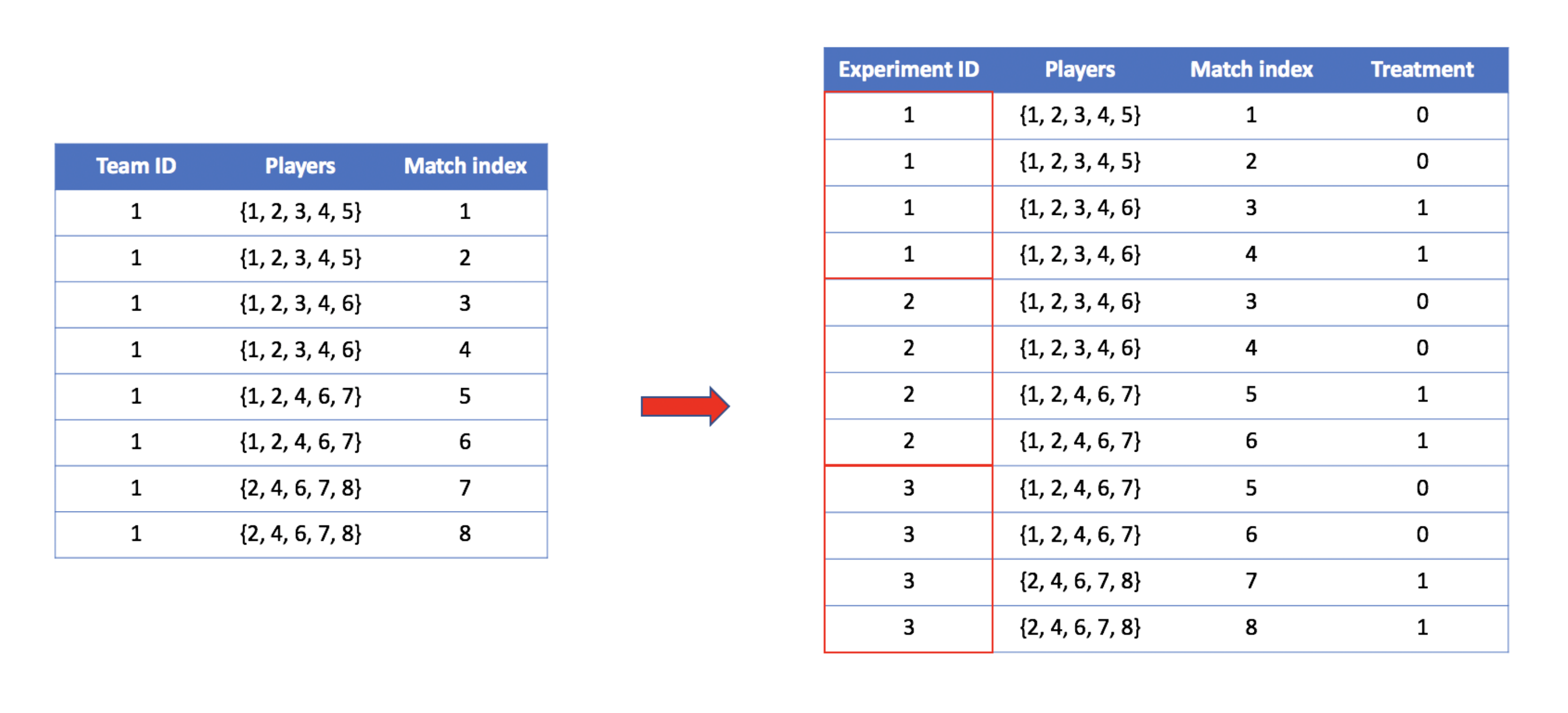}
        \caption{Teams}
        \label{fig:quasidesign_team}
    \end{subfigure}%
    \caption{Illustration of the data setup for the Customized Regression Discontinuity Design.}
    \label{fig:quasidesign}
\end{figure}

To address the specific peculiarities of our research design, we had to customize the RDD setup: 1) the event of switching team is not associated with the \textit{Match index} at the same cutoff for each player or team, and 2) players and teams are likely to have multiple times of switching. Therefore, we formulated each time of switching team as a single experiment and made the aggregate data of experiments as illustrated in Figure~\ref{fig:quasidesign}. 

In the case of players, as illustrated in Figure~\ref{fig:quasidesign_player}, each experiment consists of matches played by a player in every pair of consecutive teams. We call \textit{''Team A''} the team that a player switched \textbf{from}, and \textit{''Team B''} the team that a player switched \textbf{to}. We stacked each player's matches by the time of occurrence, and from their second time of team switching, we repeated rows of matches played in \textit{Team B} of the previous experiment to make the aggregate dataset. Likewise, in the case of teams, each experiment consists of matches played by a team with two consecutive sets of players: \textit{''Players A''} for the first match and \textit{''Players B''} for the second match. As shown in Figure~\ref{fig:quasidesign_team}, we only considered the case when the team changed one player, which means that the two sets shared four players. 

Each experiment in the two settings was assigned a unique \textit{Experiment ID}, and for each experiment, we made a new binary variable \textit{Treatment}, which had value 0 for matches played in \textit{Team A} or \textit{Players A} and 1 for \textit{Team B} or \textit{Players B}. With the aggregated data, we built \textit{Model group II} with the formulation of Eq.~\ref{eq:model2}. This model allows us to identify the effect of switching team on the baseline performance from the estimate of \textit{Treatment}. Since we controlled for the experience of the player and the team in \textit{Match index} (log-transformed to correct for the right-skewed distribution of the variable), we were able to identify the effect of switching team on the progression trend from the interaction term's estimate. The model controls for the variability in the baseline performance between experiments in the random-effects term. This model supplements \textit{Model group I} from Section~\ref{sec:cem} by estimating the change in both the baseline performance and the progression rate within every time of switching teams. 

We also controlled the prior performance in the model. For instance, when the \textit{Performance} is the number of kills, \textit{Prior performance} stands for the average number of kills the player or the team made in previous matches.

\begin{equation}
    \label{eq:model2}
    \begin{aligned}
	Performance \sim & 1 + Prior\ performance \\
	               & + Match\ index\\
	               & + Treatment \\
	               & + Match\ index \times Treatment\\
	               & + (1\vert Experiment\ ID)
	\end{aligned}
\end{equation}

To preclude the errors caused by the small sample size, we only kept the first $200$ matches of each player and each team to make the customized RDD setup. After that, we were left with 424,963 player-match combinations for 13,656 players and 152,815 team-match combinations for 6,051 teams. Additionally, to exclude cases where players played for multiple teams simultaneously, we only kept experiments where we were able to obtain at least four matches of both \textit{Team A} and \textit{Team B} for players, and of both \textit{Players A} and \textit{Players B} for teams. We were left with 10,539 experiments with 438,080 player-match observations and 1,592 experiments with 108,842 team-match observations in our final analysis. 

\section{Results}
\subsection{Number of teams and performance}\label{sec:timesst}
First, we explored the relation between the \textit{number of teams} a player played in and the player's \textit{performance}. We performed an analysis on the ''macro-level'' by comparing performance of players who switched teams often and those who rarely switched teams. To model the effect of team switching frequency, we created a coarsened exact matching design by matching players into bins by multiple exogenous variables (see \nameref{sec:matandmethods}). 

In each bin, we had players in Treatment group (who switched teams frequently) and players in Control group (who rarely or never switched teams), assigned by cut-off criteria presented in Table \ref{tab:cutoff}. We built \textit{Model group I} with the formulation of Eq. (\ref{eq:baseline_model}) that accounts for heterogeneity of bins to compare the average of each performance indicators. Please note that bins that only had observations from only Treatment or Control group were dropped ($<8\%$ of total number of observations).

\begin{figure*}
    \centering
    \makebox[\textwidth][c]{\includegraphics[width=\textwidth]{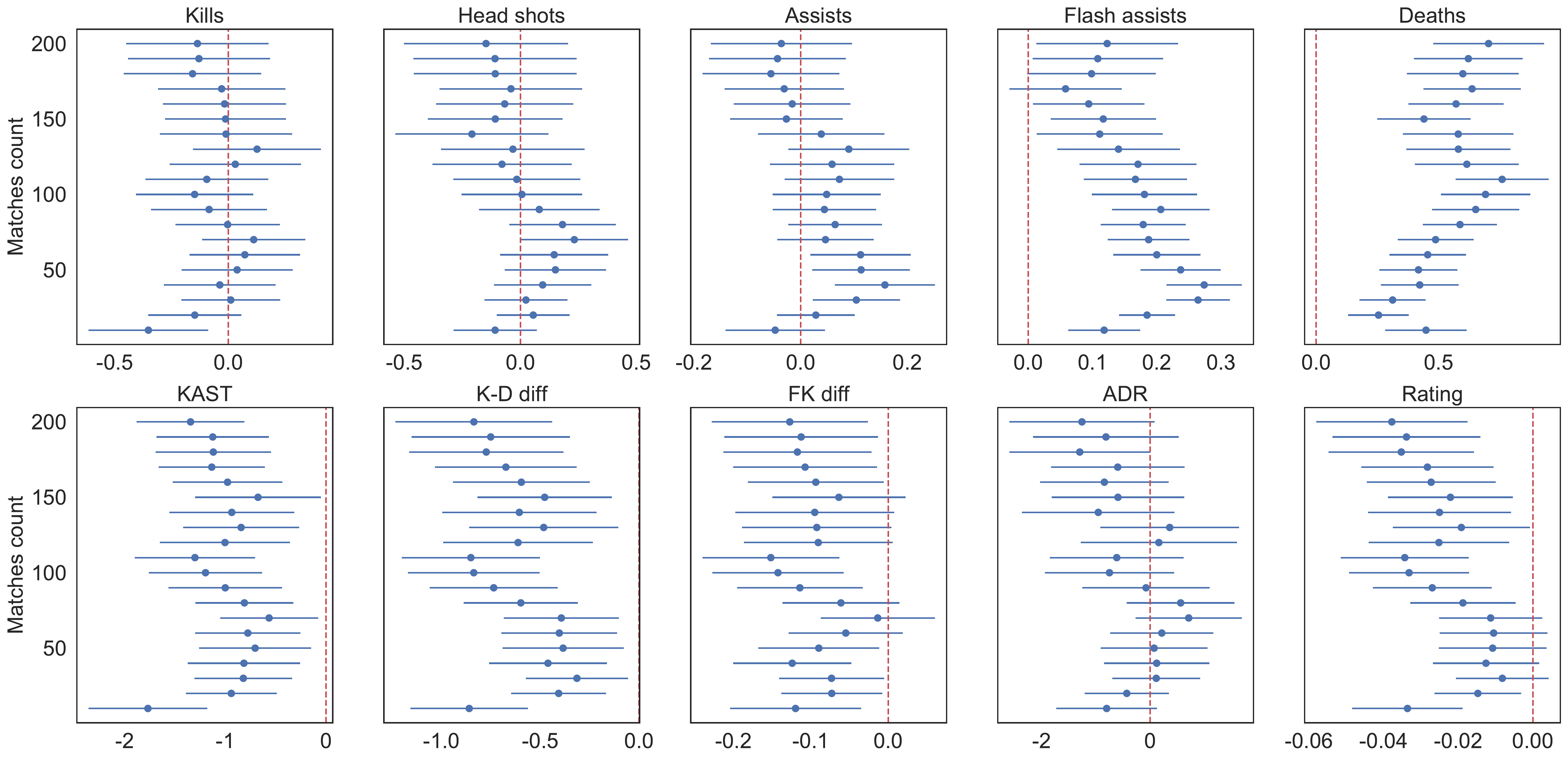}}%
    \caption{The effects of \textit{Treatment} in Model I. The points are the the mean treatment effect and the lines are the 95\% confidence intervals around the means. The effects are illustrated for the range between first 10 and first 200 matches for each player, with the increment of 10.}
    \label{fig:individual_cem}
\end{figure*}

For example, we found that the players in the Treatment group (that switch teams often) were likely to have lower performance in terms of \textit{Deaths}, \textit{KAST}, \textit{K-D diff}, \textit{FK diff} and \textit{Rating}, but perform better in making \textit{Flash assists} than those in Control group (that rarely switch teams). In the Figure~\ref{fig:individual_cem}, we plotted the coefficient of \textit{Treatment} with the 95 percent confidence interval (CI) in \textit{Model I} for different performance indicators, from the first 10 matches up to first 200 matches. The 95\% CI's show that the estimated coefficients for \textit{Treatment} in \textit{Flash assists}, \textit{Deaths}, \textit{KAST}, \textit{K-D diff}, \textit{FK diff} and \textit{Rating} are mostly significant.

The plotted coefficients in Figure~\ref{fig:individual_cem} should be interpreted in the following way. For example, the players who played at least 200 matches, those in the Treatment group roughly made 0.12 more flash assists, had 0.6 more deaths, -1.5 lower \textit{KAST}, 0.7 lower \textit{K-D diff}, 0.12 lower \textit{FK diff} and 0.04 lower \textit{Rating} on average in their first 200 matches than players in the Control groups. 

The results suggest that a player who changes teams often has lower survival rate in average, fewer contributory activities, and lower overall rating compared to those who change fewer teams. However, we did not observe a significant difference in \textit{Kills} and \textit{ADR}, which implies that switching teams frequently is not likely to affect a player's ability to cause damage to the opposing team. 

\subsection{Team switch and impact on performance}\label{sec:findrd}
We further expand the analysis from studying the total number of team changes to evaluating the immediate and long-term effects of switching the teams. We perform an analysis on the ''micro-level'' by predicting the player's and the team's performance, as well as the probability of team winning the game after each team switch. To model the effect of team-switch to the \textbf{player} and to the \textbf{team}, we employed a quasi-regression discontinuity design (see \nameref{sec:matandmethods}). 

\textbf{The effect on the player}. To estimate the effect of team switch to an individual player, we compare the performance in the matches in the team a player switched \textbf{from} (\textit{Team A}), and the matches of the team the player switched \textbf{to} (\textit{Team B}), as explained in Section~\ref{sec:rdmodel}. To predict the performance with fixed-effects terms that control information about \textit{Prior performance} and the \textit{Match index} (player's experience), we used the \textit{Model group II}, as defined in Eq.~\ref{eq:model2}.

We observed a drop in individual performance immediately after the player joins the new team. The results of the RDD analysis for individual in-game performance indicators are shown in Table \ref{tab:mod2inditeam}. Right after switching the teams, players cast less damage and show less accuracy, as indicated by the negative and significant estimates of \textit{Treatment} in \textit{Kills}, \textit{KAST}, \textit{K-D Diff}, \textit{FK Diff} and \textit{ADR}. A player who changes the team will have 0.24 less \textit{Kills} on average in the first match with a new team. The player will also have 0.22 more \textit{Deaths}, 0.61 lower \textit{KAST}, 0.30 lower \textit{K-D diff}, 0.13 lower \textit{FK Diff} and 1.04 lower \textit{ADR} than the case of not switching. However, some teamwork related aspects of the game will improve. The players become slightly more active in assisting the new teammates: the \textit{Treatment} in both the \textit{Assists} and \textit{Flash assist} models for individual performance suggests an immediate improvement.

    \begin{table*}
        \caption{Regression models that predict \textit{Individual performance} and \textit{Teammate performance} on the experiment-match level with the formation of \textit{Model group II}. $^{*} p < 0.05$, $^{**} p < 0.01$, $^{***} p < 0.001$ }
        \label{tab:mod2inditeam}
        \centering 
        \begin{center}
        \begin{adjustbox}{width=1\textwidth}
        \input{table_mod2inditeam}
        \end{adjustbox}
        \end{center}
    \end{table*}
    
We also identified some long-term changes in players' performance. The progression trend of a player becomes mildly higher after the team switch. As shown by \textit{Match index $\times$ Treatment} in Table \ref{tab:mod2inditeam}, after a player switched team, when the number of matches played increases by 100 percent, the player has 0.06 fewer \textit{Deaths}, 0.14 higher \textit{KAST}, 0.03 higher \textit{FK diff} than the scenario when the player had not switched team. And the player makes 0.04 fewer \textit{Head shots} and 0.03 fewer \text{Assists} than otherwise. These all imply that immediate drop in individuals' performance after switching teams will recover in the long run, though on a very low level. 
    
\textbf{The effect on the team}. To estimate the effect of team switch to a team, we compared the performance of all players in the team before the switch with the performance of all players in the team after the switch. The original cohort of players is referred to as \textit{Players A}, and after the new member joined, the new cohort is referred to as \textit{Players B}, as explained in Section~\ref{sec:rdmodel}.

For a team's performance, we observed similar changes in performance as in the individual player. As indicated in Table~\ref{tab:mod2inditeam} by estimates of \textit{Treatment} in \textit{Team}, immediately after a team changed one player, the team made 0.51 fewer \textit{Kills} and 0.14 fewer \textit{Head shots} and had 0.42 more \textit{Deaths}, 1.55 lower \textit{KAST}, 0.77 lower \textit{K-D diff}, 0.16 lower \textit{FK diff}, 1.32 lower \textit{ADR} and 0.038 lower \textit{Rating} in the following matches, compared to the case where the team members stayed the same. Additionally, they exhibited more teamwork activities as indicated by \textit{Assists} and \textit{Flash assists}. However, in the long run, this gap of performance are offset, as shown by \textit{Match index $\times$ Treatment} in \textit{Team}. 

\textbf{The odds of winning after team switch.} We further predicted how switching teams affect the match outcome. As shown in Table \ref{tab:mod2win}, after a team changed one player, the estimated odds of winning are 27\% lower ($\beta = -0.31$, $p < .001$), though the baseline gap of winning probability will become narrower in the long run. From a micro level, the above mentioned findings reveal that the decrease in progression rates as team affiliations increases as found in Section \ref{sec:timesst} could be attributed to the decrease in baseline performance both the player and the team. 
    
    \begin{table}
        \caption{Regression models that predict \textit{Win} on the experiment-match level with the formulation of \textit{Model group II} (N = 108,842).$^{*} p < 0.05$, $^{**} p < 0.01$, $^{***} p < 0.001$}
        \label{tab:mod2win}
        \centering 
        \begin{center}
        \begin{adjustbox}{width=.6\linewidth}
        \begin{tabular}{r l} 
        \toprule
        Intercept &-0.745 (0.04)$^{***}$\\
        Prior performance &\ 1.07 (0.04)$^{***}$\\
        Match index &\ 0.06 (0.01)$^{***}$\\
        Treatment &-0.31 (0.05)$^{***}$\\
        Match index $\times$ Treatment &\ 0.06 (0.01)$^{***}$\\
        \midrule
        Random effects\\
        \ var.Intercept &0.09\\
        Log-likelihood &-74,140\\
        \bottomrule
        \end{tabular}
        \end{adjustbox}
        \end{center}
    \end{table}
    
\section{Discussion}
The present study sought to examine how switching teams affects individual and collective performance in professional eSports team play, from both the macro and the micro-levels. Concentrating on the large-scale behavioral dataset of professional tournaments on the popular first-person shooter game {\itshape CS: GO}, we obtained a few interesting and surprising findings from the analyses. Though much research has examined team collaboration in video game playing, relatively few have focused on professional eSports athletes. We applied rigorous statistical models that revealed the team dynamics in eSports and identified the effects of switching teams on players' individual and collective performance progression. 

Before the 200$^{th}$ matches, about half of the players (N = 6,616, 48.4\%) in our dataset have played in multiple teams. Within these matches, we found that switching teams exerts a negative impact on players: the more teams a player has joined, the lower the rating the player achieves in the game in the long run. The findings suggest that unlike players who play in stable teams, players who switch teams frequently make fewer contributory actions and have lower survival rate. This could be attributed to the fact that a stable team facilitates social congruence, mutual trust and shared goals, which further benefits the group coordination and the collaborative outcome. 

We then investigated the mechanisms behind the negative impact of switching teams on the micro-level with the quasi-regression discontinuity design. After players switched to a new team, we found that players take fewer initiatives, do more occasional damage, and receive a lower rating than in the previous team. It may be that when a player changes to a new environment, it takes time for the player to get used to the group dynamic and establish effective communication with new teammates, which consequently delays their progression. 
On the other hand, for the team the player left, they also perform worse in various aspects and they become less likely to win the game than previous. These differences further suggest the lack of knowledge sharing and potential miscommunication among members in a newly formed team; or it could be because of some unhealthy competitions and distancing against the new player~\citep{lipovaya2018coordination}.

The development of mutual trust takes time in a newly formed team and a team that lacks mutual trust is likely to run into obstacles in coordination and collaboration among team members; mutual trust facilitates knowledge sharing~\citep{choi2019mechanism, pinjani2013trust} and benefits both individual functioning and team performance~\citep{de2018trust,gupta2016multilevel}. The higher the task independence, the more importance mutual trust on performance in virtual teams~\citep{de2016trust}. In the highly competitive game \textit{CS:GO}, every member is crucial for the team to win, and players have to move fast while reading signals from teammates to coordinate teamwork in the game~\citep{reeves2009experts}. This interdependence of teamwork emphasizes a stable team formation in achieving satisfying progression for both individuals and teams. 

There are incentives for proficient players to accept invitations that recruit them to play in a more competitive team that pays higher. Still, they should evaluate the benefits and costs before making the decision. As shown in our dataset, for players with available age information, 95\% of them are younger than 31 years old ({\itshape Mean} = 23.5, {\itshape SD} = 4.0). As an emerging industry, eSports require the strategies of competition to change fast~\citep{kim2015stage, meng2020understanding}. Players have to keep competing with newly joined young players, and thus their career span is short with their performance tends to slump in the later stage~\citep{salo2017career, kim2015stage, meng2020understanding}. 

From the coaching perspective, teams should also think twice when they decide to recruit new members or players who used to play in many teams, given that the change in team formation affects the new players and impacts the teammates the players play with. 

\section{Contributions and limitations}
This study provides insights for ameliorating group dynamics in eSports professional teams and informs team building strategies in computer-mediated collaborative settings. The career span of eSports players is short-lived, and we identified a performance deterioration after eSports athletes become senior players. The findings in this study may be helpful for players to optimize their career development strategies before they retire. On the one hand, collaborative team play is one primary source of enjoyment for playing video games~\citep{nardi2006strangers}, so it is critical to cultivate an agreeing team environment for players since a pleasant social experience with teammates further motivates players to keep playing~\citep{yee2006motivations}. On the other, this study also provides advice for eSports coaching by recognizing the potential harms of recruiting new members. Sponsors and supporters behind the eSports team should evaluate both the costs and benefits before hiring players who have membership in multiple teams, and after new members join, team organizers might consider initiating team building activities to enact a bonding between members. 

This study emphasizes the role of stable team formation in promoting team communication, coordination, and knowledge sharing in eSports, which enriches existing studies on computer-supported collaborative work (CSCW). Without face-to-face interactions, establishing mutual trust and knowledge sharing is more difficult in computer-mediated environments than offline organizational settings, making group communication and interactions critical to team success~\citep{pinjani2013trust}. ESports, as an emerging form of CSCW, requires intensive competitive activities, provides new opportunities and challenges to study how a healthy team environment can be formed and maintained in a virtually connected team~\citep{freeman2017esports, freeman2019understanding, lipovaya2018coordination}. By unpacking the impact of switching teams on individual and collective progression in eSports, this study points out a future research direction in CSCW.

While we generated understanding about switching teams in eSports based on a large-scale dataset, the present study did not get a chance to examine how other aspects of team composition impact the collaboration outcome. It has been found that in teams with members from multicultural backgrounds mutual trust is difficult to sustain~\citep{cheng2016investigating}. 
And based on previous studies on eSports teams, diversities in nationality and language affect team performance in different directions~\citep{parshakov2018diversity, parshakov2018determinants}. Future research could extend the present study by investigating how group composition moderates switching teams' effect on individual and collective performance. 



\newpage
\bibliography{references}


\end{document}

%% file: table_mod2inditeam.tex
\begin{tabular}{r l l l l l l l l l l} 
    \toprule
    \textbf{}&\multicolumn{2}{c}{\# Kills} 
    &\multicolumn{2}{c}{\# Head shots}
    &\multicolumn{2}{c}{\# Assists} 
    &\multicolumn{2}{c}{\# Flash assists}
    &\multicolumn{2}{c}{\# Deaths}\\
    &\multicolumn{1}{c}{Player}
    &\multicolumn{1}{c}{Team} 
    &\multicolumn{1}{c}{Player} 
    &\multicolumn{1}{c}{Team} 
    &\multicolumn{1}{c}{Player} 
    &\multicolumn{1}{c}{Team} 
    &\multicolumn{1}{c}{Player} 
    &\multicolumn{1}{c}{Team} 
    &\multicolumn{1}{c}{Player} 
    &\multicolumn{1}{c}{Team} \\
    \midrule
    Intercept 
    &11.44 (0.13)$^{***}$ 
    &13.76 (0.19)$^{***}$
    &\ 3.26 (0.05)$^{***}$ &\ 4.51 (0.01)$^{***}$
    &\ 2.22 (0.03)$^{***}$ &\ 2.01 (0.04)$^{***}$
    &\ 0.18 (0.02)$^{***}$ &\ 0.23 (0.01)$^{***}$
    &\ 13.76 (0.13)$^{***}$ &\ 14.63 (0.21)$^{***}$\\
    Prior performance 
    &\ 0.33 (0.01)$^{***}$ &\ 0.20 (0.01)$^{***}$
    &\ 0.58 (0.01)$^{***}$ &\ 0.41 (0.01)$^{***}$
    &\ 0.37 (0.01)$^{***}$ &\ 0.47 (0.01)$^{***}$
    &\ 0.65 (0.01)$^{***}$ &\ 0.69 (0.01)$^{***}$
    &\ 0.24 (0.01)$^{***}$ &\ 0.19 (0.01)$^{***}$\\
    Match index
    &\ 0.09 (0.02)$^{***}$ &\ 0.12 (0.02)$^{***}$
    &\ 0.03 (0.01)$^{***}$ &\ 0.04 (0.01)$^{***}$
    &\ 0.09 (0.01)$^{***}$ &\ 0.05 (0.01)$^{***}$
    &\ 0.04 (0.004)$^{***}$ &\ 0.02 (0.003)$^{***}$
    &-0.02 (0.01)$^{*}$ &-0.18 (0.02)$^{***}$\\
    Treatment 
    &-0.24 (0.10)$^{**}$ &-0.51 (0.10)$^{***}$
    &\ 0.11 (0.05)$^{*}$ &-0.14 (0.05)$^{**}$
    &\ 0.15 (0.03)$^{***}$ &-0.11 (0.03)$^{***}$
    &\ 0.06 (0.02)$^{*}$ &-0.03 (0.02)
    &\ 0.22 (0.01)$^{**}$ &\ 0.42 (0.01)$^{***}$\\
    Match index $\times$ 
    &\ 0.02 (0.02) &\ 0.11 (0.03)$^{***}$
    &-0.04 (0.01)$^{**}$ &\ 0.03 (0.01)$^{**}$
    &-0.03 (0.01)$^{**}$ &\ 0.02 (0.01)$^{**}$
    &-0.01 (0.01) &\ 0.01 (0.004)
    &-0.06 (0.02)$^{***}$ &-0.08 (0.03)$^{**}$\\
    \    Treatment \\
    \midrule
    Random effect
    &1.24 &0.29
    &0.56 &0.18
    &0.15 &0.09
    &0.03 &0.01
    &0.42 &0.31\\
    Log-likelihood
    &-1,426,971 &-315,608
    &-1,183,122 &-233,738
    &-993,745 &-190,654
    &-384,964 &-65,224
    &-1,305,106 &-317,353\\
    N
    &438,080 &108,842
    &438,080 &108,842
    &438,080 &108,842
    &246,986 &67,081
    &438,080 &108,842\\
    \midrule
    \textbf{}&\multicolumn{2}{c}{KAST} 
    &\multicolumn{2}{c}{K-D Diff}
    &\multicolumn{2}{c}{FK Diff} 
    &\multicolumn{2}{c}{ADR}
    &\multicolumn{2}{c}{Rating}\\
    &\multicolumn{1}{c}{Individual}
    &\multicolumn{1}{c}{Teammate} 
    &\multicolumn{1}{c}{Individual} 
    &\multicolumn{1}{c}{Teammate} 
    &\multicolumn{1}{c}{Individual} 
    &\multicolumn{1}{c}{Teammate} 
    &\multicolumn{1}{c}{Individual} 
    &\multicolumn{1}{c}{Teammate} 
    &\multicolumn{1}{c}{Individual} 
    &\multicolumn{1}{c}{Teammate} \\
    \midrule
    Intercept
    &51.55 (0.49)$^{***}$ &47.65 (0.78)$^{***}$
    &-0.45 (0.07)$^{***}$ &-0.57 (0.07)$^{***}$
    &-0.12 (0.02)$^{***}$ &-0.13 (0.02)$^{***}$
    &48.36 (0.55)$^{***}$ &49.51 (0.83)$^{***}$
    &\ 0.627 (0.007)$^{***}$ &\ 0.663 (0.011)$^{***}$\\
    Prior performance
    &\ 0.24 (0.02)$^{***}$ &\ 0.29 (0.01)$^{***}$
    &\ 0.38 (0.01)$^{***}$ &\ 0.33 (0.01)$^{***}$
    &\ 0.26 (0.01)$^{***}$ &\ 0.22 (0.01)$^{***}$
    &\ 0.36 (0.01)$^{***}$ &\ 0.33 (0.01)$^{***}$
    &\ 0.37 (0.007)$^{***}$ &\ 0.38 (0.010)$^{***}$\\
    Match index
    &\ 0.21 (0.04)$^{***}$ &\ 0.28 (0.05)$^{***}$
    &\ 0.07 (0.02)$^{***}$ &\ 0.13 (0.02)$^{***}$
    &\ 0.02 (0.01)$^{***}$ &\ 0.03 (0.005)$^{***}$
    &-0.06 (0.06) &\ 0.13 (0.05)$^{**}$
    &\ 0.006 (0.001)$^{***}$ &\ 0.007 (0.001)$^{***}$\\
    Treatment 
    &-0.61 (0.21)$^{**}$ &-1.55 (0.26)$^{***}$
    &-0.30 (0.10)$^{**}$ &-0.77 (0.11)$^{***}$
    &-0.13 (0.03)$^{***}$ &-0.16 (0.03)$^{***}$
    &-1.04 (0.32)$^{**}$ &-1.32 (0.25)$^{***}$
    &-0.007 (0.001) &-0.038 (0.006)$^{***}$\\
    Match index $\times$ 
    &\ 0.14 (0.05)$^{**}$ &\ 0.32 (0.07)$^{***}$
    &\ 0.05 (0.02) &\ 0.16 (0.03)$^{***}$
    &\ 0.03 (0.01)$^{**}$ &\ 0.03 (0.01)$^{***}$
    &\ 0.12 (0.08) &\ 0.27 (0.07)$^{***}$
    &\ 0.001 (0.001) &\ 0.008 (0.002)$^{***}$\\
    \    Treatment \\
    \midrule
    Random effects
    &3.67 &3.96
    &1.69 &0.96
    &0.12 &0.03
    &14.86 &3.66
    &0.004 &0.003\\
    Log-likelihood
    &-1,303,831 &-325,528
    &-1,479,762 &-327,440
    &-1,007,923 &-175,981
    &-1,437,002 &-322,907
    &-151,489 &-3,544\\
    N
    &331,774 &87,893
    &438,080 &108,842
    &438,080 &108,842
    &331,662 &87,876
    &438,080 &108,842\\
    \bottomrule
    \end{tabular}